# A New Approach for Finding Cloned Profiles in Online Social Networks

Fatemeh Salehi Rizi[1], Mohammad Reza Khayyambashi[2], and Morteza Yousefi Kharaji[3]

[1] Sheikh Bahaei University /Department of Computer and Information Technology, Isfahan, Iran
Email: Salehi.Fatemeh@shbu.ac.ir
[2] Isfahan University / Department of Computer Science, Isfahan, Iran
Email: M.R.Khayyambashi@eng.ui.ac.ir
[3] Mazandaran University of Science and Technology / Department of Computer Science, Mazandaran, Iran
Email: Yousefi@ustmb.ac.ir

*Abstract*—Today, Online Social Networks (OSNs) are the most popular platforms on the Internet, on which millions of users register to share personal information with their friends. Leakage of personal information is a significant concern for social network users. Besides information propagation, some new attacks on OSNs such as Identity Clone attack (ICA) have been identified. ICA attempts to create a fake online identity of a victim to fool their friends in order to reap their private information which is not shared in public profiles. There are some identity validation services that perform users' identity validation, but they are passive services and they only protect users who are informed on privacy concerns and online identity issues. This paper starts with an explanation of two types of profile cloning attacks in OSNs. Afterwards, a new approach for detecting clone identities is proposed by defining profile similarity and strength of relationship measures. According to similar attributes and strength of relationship among users which are computed in detection steps, it will be decided which profile is clone and which one is genuine by a predetermined threshold. Finally, the experimental results are presented to demonstrate the effectiveness of the proposed approach.

I. INTRODUCTION

Progress in information technology has been caused a widely change in the modality of social connections and in people relations. A few years ago, blogs, forums and instant messaging systems provided people's connections on the Internet. Nowadays, all of these mediums have been collected in social network websites which people have numerously joined them [1].

The growth of information transmission and more intimate interactions among users are the most important reasons for users to forget the negative consequences of sharing personal information on the Internet, especially when the information is shared as public data for a long time [2]. Alongside OSNs' popularity growth, security risks and threats are growing, too, which affect users' privacy and confidentiality [3]. Protecting users' privacy not only is protecting data that users share on their profiles, but also is preserving their relations and activities on OSNs [4]. An expert attacker forges identities on OSNs and tricks people into abusing their personal information [5].

In online social networks, the concept of friendship is a social link built based on the confirmation of two people who send friend's requests to each other and it is a different form of friendship in the real world, which can be used by adversaries to steal users' personal information [3]. Recently, a new type of attacks is detected that is named profile cloning or identity clone attack (ICA) that counterfeits users' profiles on social networks. The main purpose of this attack is to obtain the private information from victim's friends by forging a real profile and increasing trustworthiness in the friend circles to deceive more users in the future. Two different types of these attacks have been introduced up to now, the first one is profile cloning in the same social network, for example, when an attacker copies the users' profiles in the same social network and uses them for sending friend requests to victim's friends. An unaware victim's friend deems that the friend requests are genuine requests and unknowingly accept them, and puts his/her personal information at the disposal of the adversary; however, those are sent from the attacker. The second attack is cross-site profile cloning. The main purpose of this attack is identifying users, who have registered on a social network, but they have not registered on others. From the attacker's point of view, first target is stealing users' identities and making new profiles on the other social networks, which users still have not registered on [5].

Detecting profile cloning attacks using a more accurate approach causes to improve the active users' safety and motivate OSNs' service providers to raise the level of security and privacy of social networks services. The rest of the paper is organized as follows: In section II, attributes and friends networks on profiles are described, and some user's activities on OSNs are explained in the next section. In section IV profile cloning attacks are illustrated with more detail and a detection framework with profile similarity and the strength of relationship measure is presented in section V. All findings are experimented on a sample Facebook dataset and it is compared to previous similar works in section VI. Finally in section VII, the paper is concluded and the future works are discussed.

II. RELATED WORKS

In identity clone attacks an adversary forges the user's identity in OSNs by cheating users and misusing their reputation. Bhumiratana in [6] presented an automatic resistant model to create clone profiles of real users' profiles by applying some attack techniques in OSNs and his/her model is able to reap user's information at the specific period of time. In this model for exposing users' information more accurate, they developed previous profile cloning attacks and presented a new approach to activate fake profiles in real users' communications. Moreover, to enable this attack for detecting users' information through other OSNs, this approach maps different OSNs' activities together and uses trust mechanisms among them. Akcora, Carminati and Ferrari in [7] proposed measures to evaluate social networks users according to both their connections and profiles' attributes. They presented a measure to find users similarity without observing a big fraction of the users' network. They also proposed a measure to detect semantic similarity among users and their method is complemented by a technique to infer missing items values from users' profiles. Kontaxis, Polakis, Ioannidis and Markatos in [9] presented a tool for detecting clone profiles in LinkedIn website and their detection process included 3 steps: Information Distiller, Profile Hunter and Profile verifier. They used string matching algorithm to compare profiles' attributes, but what is being challenged is that they did not care whether attributes are public or not. They also ignored users' friend networks similarities in comparison process. Jin, Takabi and Joshi in [8] first explained all parts of a fake identity then presented a framework to discover fake identities in OSNs. They emphasized that forging an identity is not just forging the attributes of profile, and user's friend list can be counterfeit, too. To compare profile similarities, they used cosine similarity measure and their approach is based on basic profile similarity and multiple faked identities profile similarity. Although their approach can detect some cloned profiles, they do not consider the strength of relationship among profiles and do not use more accurate attribute similarity measure to compute similarity among profiles. Most of the current research has focused on protecting the privacy of an existing online profile in a given OSN. Instead, Conti, Poovendran and Secchiero in [24] noted that there is a risk of not having a profile in the last fancy social network. The risk is due to the fact that an adversary may create a fake profile to impersonate a real person on the OSN. The fake profile could be exploited to build online relationship with the friends of victim of identity theft, with the final target of stealing personal information of the victim, via interacting online with the friends of the victim.

III. PROFILES IN SOCIAL NETWORKS

Although social networks sites have many differences, there is a common concept among them such that each user presents a unique entity to show him and has a changeable profile, which uses to demonstrate his/her image and list of attributes to anyone in online world. Social networks websites provide a general template for users' attributes and allow them to enter arbitrary activities, interests, music, movie, and general information about him, and users also may add arbitrary fields to his/her profile [10]. Most profiles include two parts: attributes fields and list of friends. As an example, user's profile items are demonstrated in Fig. 1 Items on social network profiles can assort according to two vertical dimensions. First classification is based on single value (e.g., gender) and multiple values (e.g., hobby), the second dimension is related to weather they have several subfields (e.g., education) or not. In Fig. 1, single value items (gender) are shown with multiple value items (hobby) as well as multiple value items with subfields (education and work). Values of these items are shown with their numbers like 1, 2, and 3. Work and education items have two subfields (e.g., school) while hobby has one subfield for each value [7]. Moreover, Users activities on social networks are: Like [13], Comment, Private Messages.

IV. PROFILE CLONING

One of the serious concerns along growing OSNs is increasing identity theft attacks which an adversary steal user's personal information and other related data and use them to fulfill his/her objectives. One of the most reasons of these attacks is growing web applications which use users' personal data and can be target for adversaries easily. An identity thief is a person who attempts to steal users' identities illegally and use them for tricks and financial abusing. Although people often deem that an identity theft is a strange person, statistics show that the adversary already knows victim and he might be one of victim's relatives, family members, friends and colleagues [14].
Profile cloning attack (also called identity clone attack) attempts to create a fake identity of victim in OSNs to trick their friends in to believing the validity of the fake identity, to make social links, and catch private information of victim's friend successfully which is not shared in their public profiles [6]. Bilge, Strufe, Balzarotti and Kirda in [5] introduced two types of these attacks in online social network sites. In the first attack, an already existing profile on social network is cloned and several friend requests are sent to victim's contacts. Thus, an adversary steals user's contacts by forging his/her identity and making another similar or same profile in the same social network. By obtaining victim's contact, accessing the sensitive personal information of these contacts is possible. Experimental results showed that a typical user would like to accept a friend request of a fake identity that is a confirmed contact in their friend list. In the second attack, users who are registered on a social network but they have not registered on others are identified automatically. Victim identity in the website where he/she has registered is cloned, and it is counterfeited in other social networking site where he/she has not registered yet. After creating a fake profile successfully, friend network of victim is reconstructed by contacting his/her friends who registered on both social networking sites [15]. Experimental results demonstrated this attack is very effective because profiles in this case are available just on one social network site that is being aimed. So that a friend request which sent seems completely legitimate and do not enhance suspicious of users who have been contacted [5]. Fig. 3 shows how these attacks are happen in OSNs, the left picture presents the attack which cloned the available user's account within the same social network site and the right one shows the attack which cloned users' account in other social network site.

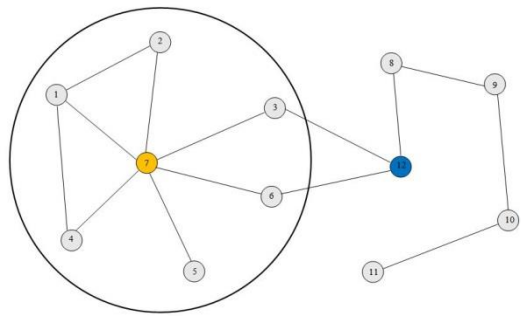

Figure 1. An example of a profile for user v. The profile consists of single valued gender and hometown items, as well as multiple valued education and work items [7]

Figure 2. Friend network graph of user 7 in a social network

In the second, more advanced attack, cross- site profile cloning attack that victim's contacts are reestablished in another social network where victim is not registered yet [15]. In this paper, attributes and friend network are used as the main target by the adversary to forge users' attributes and rebuild victim's friend networks to raise trust in friend circle [5]. In the most OSNs, name is a public attribute and it treat as a keyword of identity information. Consequently, as the first step of ICA, an adversary often builds a fake identity which has a same or similar name with victim, besides name, he might forge several other attributes which are similar with victim. Victim's attributes may find from his/her homepage or public profile in other OSNs but Sometimes an adversary cannot access this information easily while a user does not have a homepage or he/she has a lot of care to reveal his/her information publicly. However, although victim's public information is not more enough, it cannot prevent ICAs in which an adversary plans accurately. He can set attributes as the private attributes when he does not know value of victim's attributes. Forasmuch as users do not connect fake identities, they just see public attributes and they may not distinguish between fake and genuine profile. An expert adversary is able to guess victim's privacy setting by checking empty field of public profiles. By doing this, an adversary can forge not only the value of attributes that he/she does not know, but also conceal some information that is regarded as private and they are set as public by victim so that these activities may show fake identities more genuine [8].

The main goal of ICA is adding victim's friends and makes a more trustworthy relationship with them. Afterward, an adversary builds a fake identity based on similar attributes with victim and the next stage is forging victim's friend network. First an adversary might try to gain friend list from victim's profile then he send friend requests to all of these friends to forge victim. Indeed experimental results showed that a sample user tends to accept friend request from fake identity, even its real counterpart exist in his/her friend list [5]. An adversary may add many victim's friends successfully in his/her friend network. When it gets sufficient friends, it can forge victim more accurate [8].

V. PROFILE CLONING DETECTION

In order to detect cloned profiles in a same social network, the following steps are proposed:

*A. Collecting Suspicious Profiles*

In this stage, some attributes such as first name, family name, location are extracted from a real profile (the profile owner willing to find his/her clones). This information is used for making test queries in search engines available on OSNs' search service. The search results are a list of suspicious profiles which are used in next stage.

*B. Profile Evaluation*

In this stage all of suspicious identities are collected in previous stage and they are evaluated based on attributes similarity and strength of relationship measures compare to real profile. When amount of profile similarity is more than predefined threshold and strength of relationship is less than others it is determined as a clone profile which is made by an adversary and it is put in fake identity list. In the other hand, when an identity specify as a real the amount of trust is increased to avoid more validation in future. Eventually, fake identity in fake list are deleted or closed temporary and their friends are received some notifications from social network service provider or profile owner.

*C. Attribute Similarity Measure*

Attribute similarity measure calculates attribute similarity between two profiles based on values in their fields. For defining similarity measure, it is considered that users' profiles often include categorical data [16]. If two items have the same value, the most basic measures to compare two attributes return1; otherwise, they return 0 [7]. As mentioned in section III, items on a social network profile can be categorized two orthogonal dimensions; they are single valued or multiple valued. The second dimension is related to whether they contain multiple subfields or not. Similarity for single value items is computed as follows:

*C.1 Definition 1 (Single value similarity)*

Let i be a single value profile item consisting of set of subfields Sb. Let $i_v$ and $i_c$ be the value of the item i in v and c profiles, respectively. The value of n-th subfield for $i_v$ is denote as $i_v^{(n)}$. Single value similarity is giving by the following:

$$Sim_{sv}(i_v, i_c) = \frac{1}{|Sb|} \times \sum_{n \in Sb} \begin{cases} 1 & if\ i_v^{(n)} = i_c^{(n)} \\ 0 & if\ i_v^{(n)} \neq i_c^{(n)} \end{cases} \qquad (1)$$

Single value similarity computes the similarity of two single-values on different profiles. Some items have more than one value and therefore similarity must be totalized for computing item similarity. In this summation, first single-value similarity for each pair of item values on two profiles is computed then the highest one is chosen to compute item similarity. For each item, Item similarity is computed as follow:

*C.2 Definition 2 (Item Similarity)*

Let i be a profile item consisting of s set of subfields Sb. Let values ($i_v$), values ($i_c$) be the set of values for item i in u and x profiles, respectively. $\forall$ h ∈ values ($i_v$), let SimItem (h) = {$Sim_{sv}$(h, k) | k ∈ values ($i_c$)} be the set of single value item similarity computed between an element h in values($i_v$), and elements in values ($i_c$) . The similarity for item I is defined as:

$$Sim_{iv}(v,c) = \frac{1}{|values(i_v)|} \times \sum_{h \in values(i_v)} max(SimItem(h)) \qquad (2)$$

Although item similarity calculates similarity of items on two profiles, social network profiles do not include only single items. In fact, most social networks have tens of profile items, such as education and home town [7]. For finding profile similarity, these item similarities must be totalized. Some items have different weights to calculate the similarity, thus, a weighted model is applied in summation. It allows indicating importance of some items over others. Hence, the profile similarity of profile *v* and profile c is defined as follows:

*C. 3 Definition 3 (Profile Similarity)*

For $v, c \in G$:

$$Sim_{prf}(v,c) = \frac{1}{|I|} \sum_{i \in I} \beta_i \times Sim_{iv}(v,c) \qquad (3)$$

Where $\beta_i$ and $Sim_{iv}(v,c)$ are important and the similarity of the ith item respectively, and I is the item set on user's profile. Importance coefficients are user- defined [7].

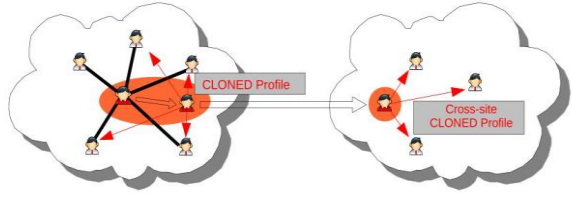

Figure 3. Profile cloning and Cross-site profile cloning in online social networks [15]

*C. 4 Example of attribute similarity calculation*

Given a scenario where values of the education item of profile v and profile c are given. For instance, user v and user c have two education values on their profiles and also each value has two subfields as is shown in Fig. 4.

Given the information on two profile items, similarity measure of *v*'s education value and *c*'s education value is computed in the following.

There are two subfields (school and degree) in education value thus, $Sb = 2$ and single- value similarities for all pairs of item values on profile *v* and *c* are computed as follows:

$$Sim_{sv}(i_v, i_c) = \frac{1}{|2|} \times (1+1) = 1, \qquad Sim_{sv}(i_v, i_c) = \frac{1}{|2|} \times (0+0) = 0$$

$$Sim_{sv}(i_v, i_c) = \frac{1}{|2|} \times (0+0) = 0, \qquad max(SimItem(h)) = 1$$

Next item similarity should be computed as follows:

$$Sim_{iv}(v,c) = \frac{1}{|2|} \times (1+0) = \frac{1}{2}$$

Finally, after computing all of item similarities, profile similarity should be computed by (3), but in Example1, it is just assumed to compute until item similarity stage.

*D. Strength of relationship measure*

In this section, social networks data are modeled using a weighted graph, such that user profiles are represented as nodes and their interactions are shown as edges in the social graph. It should be considered that the linkage between profiles are used to model user interactions, rather than their friendship relations.

Formally, the social network is defined as a weighted graph $G = (V, E, W)$, where $V$ is the set of profiles, $E \subseteq V \times V$ is the set of edges, and $W \subseteq \mathcal{R}$ is a set of weights are assigned to edges. For each node $v \in V$, a 3-dimentional feature vector is defined as it is included in the number of active friends, page likes and common shared URLs.

Therefore, weight of each edge $e_{ij} = (v_i, v_j)$ is calculated as summation of common actives friends, page likes and common shared URLs between nodes $v_i$ and $v_j$. Further details presented how the weights can compute come in the following parts.

*D.1 Active friends*

This measure takes the interaction frequency of a user with his/her friends in the network. For a user $V_i$ with $F_i$ as the set of friends, the set of active friends $F_i^a$ can be computed as an interaction between the set $F_i$ and the set of friends of $V_i$ who were either contacted by $V_i$ or those who interacted with $V_i$ through wall posts, comments or tags. It can be defined using (4), in which $I_i$ is the set of users with whom $V_i$ has interactions in the network. For a node $V_i$ the value of the "active friends" feature is taken as the cardinality of the set of its active friends $F_i^a$. Similarly, the set of common active friends in the network with whom a pair of users $v_i$ and $v_j$ have interacted is calculated as the intersection of their active friends $F_i^a$ and $F_j^a$, respectively, as given in (5). for an edge $e_{ij} = (v_i, v_j)$, the value of the "active friends" feature is taken as the cardinality of the set of common active friends $F_{ij}^a$ [17].

$$F_i^a = F_i \cap I_i \qquad (4)$$

$$F_{ij}^a = F_i^a \cap F_j^a \qquad (5)$$

*D.2 Page Like*

This feature computes the page likes frequency of the users in social network. For an edge $e_{ij} = (v_i, v_j)$, the common page likes of $v_i$ and $v_j$, $P_{ij}$, is calculated as the interaction of the sets of page likes of $v_i$ and $v_j$, as given in (6), and the page likes attribute value is calculated as the cardinality of the set $P_{ij}$ [17].

$$P_{ij} = P_i \cap P_j \qquad (6)$$

*D.3 URLs*

This feature captures the URL sharing patterns of the social networks users. For an edge $e_{ij} = (v_i, v_j)$, the common URLs of $v_i$ and $v_j$, $U_{ij}$, is calculated as the intersection of the set of URLs shared by $v_i$ and $v_j$. The URLs attribute value is calculated as a fraction of URLs commonly shared by them using (7) [17].

$$U_{ij} = \frac{U_i \cap U_j}{U_i \cup U_j} \qquad (7)$$

Based on above mentioned features, each edge $e_{ij} = (v_i, v_j)$, is assigned a weight $w(e_{ij})$ that is calculated as a summation of the individual feature value as given in (8). | | represents the cardinality of the set [17].

$$w(e_{ij}) = |F_{ij}^a| + |P_{ij}| + |U_{ij}| \qquad (8)$$

Afterward, the weights are assigned to each edge in social network graph that is shown in Fig. 5, strength of relationship should be calculated between two nodes as follows:

*D. 4 Definition 4 (Friendship Graph)*

Given a social network G and a node $v \in G.N$, the friendship graph of v, denoted as FG (v), is a sub-graph of G where: (1) FG (v).N = {v} ∪ {n ∈ G.N | n ≠ v, ∃ e ∈ G.E, e = <v, n> }; (2) FG (v).E = {e = <v, n> ∈ G.E | n ∈ FG (v).N} ∪ {e = <n, n′> ∈ G.E | n, n′ ∈ FG (v).N } [7]

*D. 5 Definition 5 (Mutual Friends Graph)*

31

Given a social network G and two nodes v, c ∈ G.N, the mutual friends Graph of v and c, denoted as MFG (v, c), is a sub-graph of G where: (1) MFG (v, c).N = {v, c} ∪ {n ∈ G.N | n ≠ v, n ≠ c, ∃ e, e′ ∈ G.E, e = <v, n> ∧ e = <n, c> }; (2) MFG (v, c).E = {e = n, n′ ∈ G.E | n, n′ ∈ MFG (v, c).N} [7]

*D. 6 Definition 6 (Strength of relationship between two nodes)*

Given a social network G and two nodes v, c ∈ G.N, Let T = {MFG (v, c).E}, R = {FG (v).E}, P = {FG (c).E}. Strength of relationship between v and c is defined as follows:

$$Rs\ (v,c) = \frac{\sum_{r \in T} w_r}{\sum_{r_1 \in R} w_{r_1} + \sum_{r_2 \in p} w_{r_2}} \quad (9)$$

Strength of relationship measure is calculated between suspicious profiles which have common friend with victim. Inasmuch as an expert adversary attempts to make less suspicious by making relation and interactions with victim's friends thus, strength of relationship measure can be used to detect profile cloning. An adversary try to indicate his/her fake profile more real by having common friends and doing different activities such as like, share URLs, etc. Hence, using strength of relationship measure between suspicious nodes and real nodes it will be found out which profile is clone with high probability, because the fake identities cannot interact as much as the real identities can connect to other users. A node that has more attribute similarity and similar activities with a real node makes less suspicious for a simple user.

*E. Discovering communities in social network*

Most social networks users post their interest topics on their walls and invite their friends to express their ideas. Hereby, they provide a rich source of data for studying users' relationships and interaction patterns on large scale.

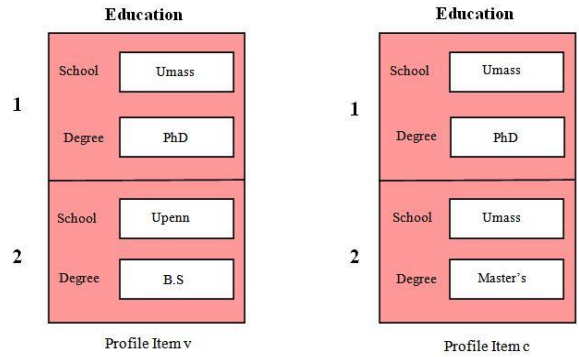

Figure 4. Education items on profile *v* and profile *c*

A community is a set of users as a group with high relatedness among people inside the group. Collection of people into communities gives important insights into group dynamics which can be used for online marketing, information dissemination and understanding the formation of clubs, committees and action groups in real world. Recently a generative Bayesian model to discover topic based community was proposed to extract latent community from social network graph [18]. A smart adversary creates a clone identity in which social network acts in the same community with real identity. It tries to seem more genuine and it participates in same interest topics with real identity. Accordingly, a candidate identity who is more similar to real node and has less strength of relationship measure within real node's community would be a clone identity with high possibility.

*F. Evaluation and detection*

A detection process is designed for identifying fake profiles on OSN sites based on the profile similarity measure and strength of relationship measure as shown in Fig. 6. Given an input identity and a profile set, all of the identity profiles that have similar Name and Gender are collected by search service. After obtaining candidate profiles, profile similarity and strength of relationship measure between each candidate identity and real identity are calculated based on the schemes that are suggested in this paper. When the profile similarity of a candidate identity is larger than a predefined threshold and strength of relationship measure is less than others it is added in to suspicious identities. Input identity is also added to suspicious identities list inasmuch as it is not known whether input identity itself is authentic or not. Every identity in suspicious identities list is validated in this step such that when an identity is verified to be a fake one, it is regarded as fake identities. Then again, when an identity is verified to be a genuine, its trust value is raised for avoiding future multiple validations. Finally the clone identities are temporally closed or deleted and exiting friends of these identities will be received notifications that some of their friends are determined as fake identities.

Here, several approaches for evaluating identities which are determined as a suspicious are discussed. A primitive approach to identify users is asking them to enter their real ID. For example Identity Badge needs users to enter their passport to verify them [19].

However, these types of approaches may be difficult for users to use it in the world. Furthermore, inasmuch user's unique ID in real weird may be hard to change, it is difficult for victim to recover from an ICA when an adversary knows victim's unique ID and uses it to pass identity validation in OSN sites. Other approaches that recently have used to identify users are presented by MysafeFriend [20], which is a third-party application of Facebook. It sets five levels of trust for each identity that starts from low levels (level 1 to level 3) and it asks an identity to select his/her friend for verifying him. Howsoever more friends verify this identity, it is received more points and eventually it enhances to high levels. In the high levels, the application checks identity's credit card to verify them. Though, asking user's friends to verify directly weather an identity is real or not, that is not secure enough. It is better to ask his/her friends to design some questions for a user and evaluate the solutions for him/her. Verifying an identity should be based on the number of questions that he could answer correctly. In addition, the questions which ask users to input their personal information like credit card may not be effective for users who likely avoid them.

A Social validation was subjected by Schecher, Egelman, and Reeder [21] presented an approach while the user answer the most friends' questions correctly he/she is verified as valid user. This approach is ideal because it does not request any private information for validation. However, the current approaches are not safe completely.

There are some issues in this area, for instance, it is needed a mechanism to choose appropriate friend's identity to design questions for identifying them, and also how this questions should be designed to verify identities more effective.

This approach should not be shown identity's all of private data because it is possible for a validation service to choose a fake identity for verifying other identities. Another approach is monitoring and analyzing fake identity's activities such as login times, and check friend times, though it is not guaranteed that it works until a practical research is done. Finally, validation process might verify double identities of a user as a fake identity, when user avoids answering validation process for his/her multiple identities. In addition, some real identities may fail during validation process such that they do not answer validation questions in the specific period of time. As a result, a request might be applied for identities that are determined as fake mistakenly.

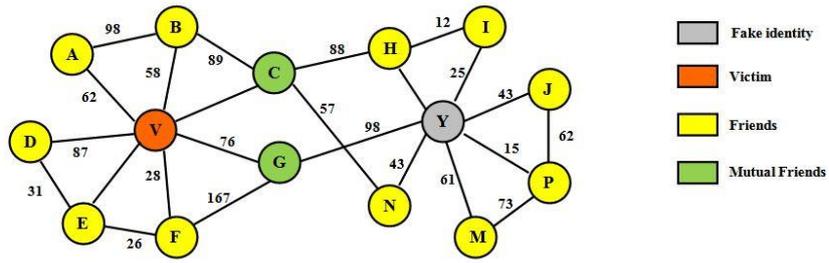

Figure 5. A weighted friend network graph in a social network

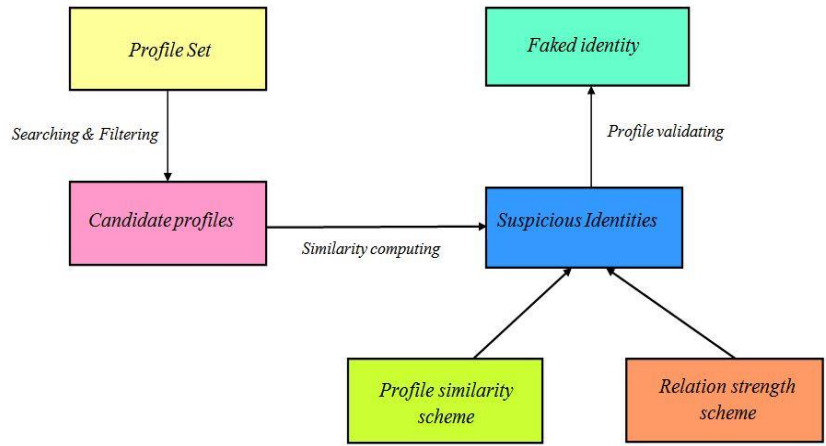

Figure 6. The detection process

VI. EXPEIMENTAL RESULTS

In this section, the experimental results are presented to validate detection model. The offline dataset of Facebook [22] is used and users' attributes, page likes and shared URLs is added them to evaluate detection process. There are 63,731 users in this dataset and 1,634,115 links among them thus each user has 25.6 relationship links on average. The detection framework cannot be validated on real system because it is difficult to find verified identities and their suspicious identities on OSNs sites. Also, faked identities cannot be created into real system as such activity interferes with OSN sites [5]. In fact, a special set of identities is assumed as fake identities and they are added as red nodes in social network graph. Most users have active friends who comment and write on their wall and they are obtained for each user by Facebook wall post dataset [23]. Due to detect clone profiles in proposed approach precisely, communities should be discovered in social network graph. Proposed detection model are implemented on the nodes of social graph and after that it will be compared with previous detection models.

34

For doing this, fist two attributes name and gender are extracted from real users' profiles then similar profiles are sought among identity profiles in dataset and profiles which have same gender and similar name are collected. Afterwards, profile similarity measure between each candidate identity and real profile are computed based on attribute similarity schemes in regard to (1), (2), (3). Strength of relationship between each candidate identity and real profile must be computed with respect to (9). Here, according to predefined threshold (δ = 0.005) and the profiles with high similarity and less strength of relationship (minimum RS) with same community of victim are selected to more evaluation by mutual friends.

In order to demonstrate the accuracy of the new approach, first two parameters must be defined as follows:

True positive (TP): Number of clone nodes that are identified as fake nodes

False Positive (FP): Number of real nodes that are identified as fake nodes

Next, some other clone nodes are added to dataset in each step (100, 200, 300, 400 clone nodes) and new approach is applied on. As shown in Fig. 7, for all numbers of fake nodes, the mount of TP is higher than FP.

With the intension of comparing new approach to previous approaches, all of three previous approaches are applied on the dataset. As diagram in Fig. 8 shows, in previous approaches the mount of their TP is less than the TP of new approach and also the mount of their FP is more than the FP of new approach. Hence our approach can detect fake nodes more accurate than others.

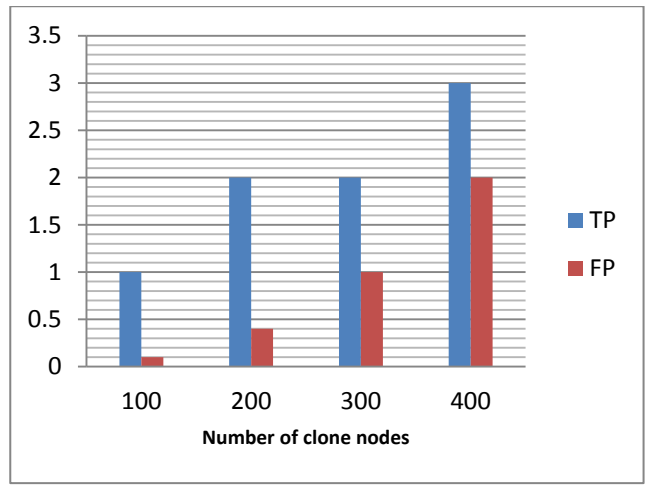

Figure 7. TP and FP for clone node detection

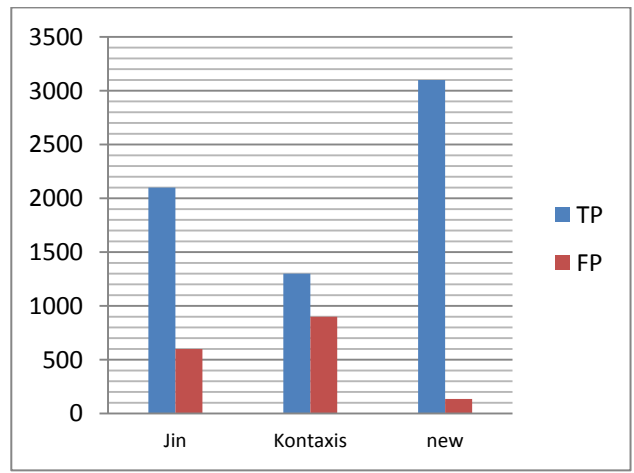

Figure 8. Comparing three exiting approaches

VII. CONCLUSION

Today, social networking sites have become popular among users in all ages, and majority of activities which users do on the Internet are allocated to these sites [25]. Many business and personal interactions are being done on social networks and meantime profile cloning becomes a serious threat in OSNs. When the identity theft happens successfully, the attacker makes friendship links with victim's friends and accesses to their personal information [5]. In this paper to detect cloned identities, two detection schemes have proposed. According to profile similarity and strength of relationship measures, it will be decided which profile is clone of real identity and which one is genuine. Through experiments, it was shown that the presented approach is very effective and it can discover clone identity more accurate. As future work, Facebook or LinkedIn third party application could be developed to implement the proposed detection schemes in real environment. Also, a new mechanism for user verification could be designed to add social networks especially when users send friend requests to each other in order to have more real friends in online social network environments.